\documentclass[12pt]{article}

\usepackage{epsfig}
\newcommand{\beq}{\begin{equation}}
\newcommand{\eeq}{\end{equation}}
\newcommand{\beqa}{\begin{eqnarray}}
\newcommand{\eeqa}{\end{eqnarray}}
\newcommand{\beqar}{\begin{eqnarray*}}
\newcommand{\eeqar}{\end{eqnarray*}}

\begin{document}

\thispagestyle{empty}

\hfill{}

\hfill{}

\hfill{}

\hfill{hep-th/0304124}

\vspace{32pt}

\begin{center}
\textbf{\Large A note on accelerating cosmologies from compactifications
and S-branes
}

\vspace{40pt}

Roberto Emparan$^{a,b}$ and
Jaume Garriga$^a$

\vspace{12pt} $^a$\textit{Departament de F{\'\i}sica Fonamental}\\
\textit{and C.E.R. en Astrof\'{\i}sica,
F\'{\i}sica de Part\'{\i}cules i Cosmologia,}\\
\textit{Universitat de
Barcelona, Mart\'{\i} i Franqu\`es, 1, E-08028, Barcelona, Spain}\\
\vspace{6pt}
$^b$\textit{Instituci\'o Catalana de Recerca i Estudis Avan\c cats (ICREA)}\\
\vspace{6pt}
\texttt{emparan@ffn.ub.es, garriga@ifae.es}
\end{center}

\vspace{40pt}

\begin{abstract}

We give a simple interpretation of the recent solutions for cosmologies
with a transient accelerating phase obtained from compactification in
hyperbolic manifolds, or from S-brane solutions of string/M-theory. In
the four-dimensional picture, these solutions correspond to bouncing the
radion field off its exponential potential. Acceleration occurs at the
turning point, when the radion stops and the potential energy
momentarily dominates. The virtues and limitations of these approaches
become quite transparent in this interpretation.

\end{abstract}

\setcounter{footnote}{0}

\newpage

There is a current effort to understand the recently discovered cosmic
acceleration, as well as to embed the inflation paradigm, within a
fundamental framework such as string/M-theory. Progress in this
direction has been hampered by some generic arguments, which are often
regarded as a no-go theorem, that forbid cosmic acceleration in the
cosmological solutions obtained from compactifications of any pure
supergravity model \cite{nogo}. In more detail, one assumes that the internal
manifold is compact, static, and non-singular.

In recent work, Townsend and Wohlfarth have pointed out that it is
possible to find solutions that exhibit a transient phase of
acceleration if the size of the internal manifold is allowed to change
in time \cite{tw}. Further extensions of this idea have appeared more
recently in \cite{ohta2,wolf,roy}, where it is argued that a similar
behavior is present as well in certain time-dependent solutions of
supergravity with $p$-form fluxes, known as S(pacelike)-branes
\cite{cgg,ohta1,sbunch}. Earlier work on closely related cosmologies can
be found in \cite{low,cc}.

Our purpose in this note is to point out that the qualitative properties
of these solutions become particularly transparent by going to a reduced
four-dimensional description. In this framework, the solutions are viewed
as cosmologies with scalar exponential potentials, and the origin of the
accelerating phase admits a simple and illuminating interpretation.

The essence of the idea is as follows. The construction of these
models involves compactification on a maximally symmetric space,
plus possibly the fluxes of covariantly constant four-form field
strengths. In the reduced description, all of these give rise to
exponential potential terms for the compactification volume scalar
$\psi$, also known as the ``radion". For sufficiently large
negative $\psi$, the potential is a positive exponential
$V(\psi)\sim +e^{-\alpha\psi}$, while for $\psi\to +\infty$, the
potential goes to zero as $V(\psi)\sim \pm e^{-\beta\psi}$.
Suppose now that the field starts at a large value of $\psi >0$,
with very large negative velocity $\dot\psi$. This corresponds to
a kinetic-dominated regime with equation of state $p\simeq\rho$.
So the universe starts out in a phase of decelerated expansion. If
the initial kinetic energy of $\psi$ is large enough, then at a
later time we will find the field running up the exponential
potential hill $+e^{-\alpha\psi}$. Eventually it will stop, due to
the increase in potential energy and to cosmic friction. So around
this point, the potential energy dominates: the universe enters an
accelerating phase. Soon after this, the field starts rolling back
down the hill, gaining kinetic energy again and entering a new
decelerating phase. The endpoint of the evolution depends on the
details of the potential --- typically the field rolls back down
to $\psi\to\infty$, but depending on whether the potential is
asymptotically positive or negative, there will be a slowly
decelerating expansion, or collapse. These are the attractor
solutions of the system.

This kind of behavior is precisely the one exhibited by the
solutions of \cite{tw}. It is then clear that the acceleration is
all the result of initial conditions where the field starts out
with a large kinetic energy, so it can climb up the positive
exponential potential and attain a brief interval of potential
domination around the turning point. The expansion that one can
get in this way seems rather modest. On the other hand, the
exponential potentials that arise from either compactification or
four-form fluxes are not enough to yield power-law inflation.
Therefore it seems just as difficult as before to obtain a
sufficient number of e-foldings of accelerated expansion in this
class of cosmologies.

Let us now flesh out the previous discussion with some calculations.
We take our higher-dimensional Lagrangian to be
\beq\label{hidlagr}
I=\frac{1}{16\pi G_{4+n}}\int d^{4+n}x \sqrt{-G} \left( R[G]
-\frac{1}{2\times 4!} F_{[4]}^2
\right)\,,
\eeq
and let the geometry be a warped product of a four-dimensional spacetime
and an internal compact space $\Sigma_{\sigma,n}$ which we shall assume to be
maximally symmetric, with $R_{ab}(\Sigma_{\sigma,n})=\sigma
(n-1)g_{ab}$, $\sigma=-
1,0,+1$,
\beq\label{compac}
ds^2=e^{-n\psi(x)} g_{\mu\nu}(x)dx^\mu
dx^\nu+e^{2\psi(x)}d\Sigma^2_{\sigma,n}\,.
\eeq
The field strength is taken as $\ast F_{[4]}= b \,{\rm
vol}(\Sigma_{\sigma,n})$ (we consider $b>0$ without loss of generality).
The factor $e^{-n\psi(x)}$ in front of the four-metric $g_{\mu\nu}$ has
been chosen so that $g_{\mu\nu}$ is the Einstein metric in four
dimensions. Indeed, upon the dimensional reduction we get \cite{gv}
\beq\label{lodlagr}
I=\frac{1}{16\pi G_{4}}\int d^4x \sqrt{-g}\left( R[g]
-\frac{n(n+2)}{2}(\partial\psi)^2 -
V(\psi)\right)\,,
\eeq
where
\beq\label{psipot}
V(\psi)= \frac{b^2}{2}\; e^{-3n\psi}-\sigma
n(n-1)e^{-(n+2)\psi}\,.
\eeq
A case of particular interest is $n=7$ since it corresponds to a
compactification of eleven-dimensional supergravity on
$\Sigma_{\sigma,7}$. For a vacuum solution, $b=0$, as considered in
\cite{tw}, we see that in order to have a positive potential the compact
manifold has to be hyperbolic, $\sigma=-1$ \cite{hyperb}. Allowing for a
non-vanishing flux $b$ always gives a positive exponential term.

\begin{figure}[ht]
\begin{center}\leavevmode  %
\epsfxsize=16.5cm\epsfbox{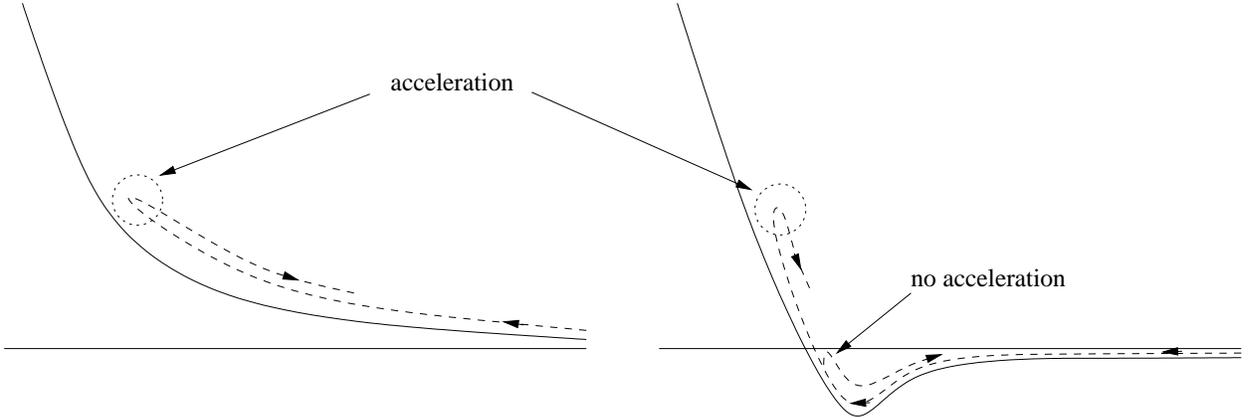}
\end{center}
\caption{Motion of the radion $\psi$ in the potential $V(\psi)$.
In the left figure all the exponentials in the potential are positive.
The field starts out at $\psi\to+\infty$ with very large (infinite)
kinetic energy. Around the point where it turns around, the energy is
potential-dominated and acceleration occurs. In the right figure there
is a combination of a positive exponential (coming from four-form flux)
and a negative exponential (from spherical compactification). We show
two trajectories: one reaches $V(\psi)>0$ and hence produces positive
acceleration (although the universe may be expanding or collapsing). The
other turns around at $V=0$ and therefore does not lead to acceleration.
All these possibilities are realized in the solutions described in the
text. The asymptotic future behavior is determined by the attractor
solutions of each potential.
}
\label{fig:exppot}
\end{figure}

The potential (\ref{psipot}) is of the sort we described in the
introduction. The different possibilities are plotted in figure
\ref{fig:exppot}. One can now use the intuition gained from the study of
cosmologies driven by scalar potentials to understand qualitatively the
possibilities permitted by time-dependent compactifications of
(\ref{hidlagr}). For example, all cases with positive exponential
potentials will be qualitatively quite similar. So hyperbolic
compactifications, with or without fluxes, and flat compactifications
with fluxes, will exhibit similar behavior, and all of them will admit a
phase of accelerated expansion. Details such as the rate of acceleration
can change depending on where in the potential the field comes to a
halt. Also, the late time behavior will be different, since the
attractor depends on the sign and exponent of the dominating exponential
term for large $\psi$. But it is easy to infer the qualitative features
of the solution from the general properties of $V(\psi)$.

The case that introduces more complications is that of spherical
compactification. If $b=0$ the potential is a negative
exponential, and clearly there is no possibility of an
accelerating phase. When $b\neq 0$ there is a competition between
a positive and a negative exponential, and there appears a wider
variety of possible behaviors. However, acceleration is clearly
possible for solutions where $\psi$ has a turning point at
positive values of $V(\psi)$.

We can check the correctness of this qualitative discussion using
the time-dependent solutions of the Lagrangian (\ref{hidlagr})
found in \cite{cgg}\footnote{It is possible to consider as well a
dilaton $\phi$, coupling as $e^{-a\phi} F_{[4]}^2$. This has the
virtue that it allows to recover the solutions of \cite{tw} from
the ones in \cite{cgg} with $\sigma=-1$ by simply letting $a\to
\infty$. As we explained, all hyperbolic compactifications, with
our without flux, are qualitatively similar.}. The ansatz used
there for the solutions of relevance to us is \beq
ds^2=-e^{6B+2nC}dt^2 +e^{2B}d{\bf
x}_{(3)}^2+e^{2C}d\Sigma^2_{\sigma,n}\,, \eeq with $B$ and $C$
functions of $t$ only. We can translate the results of \cite{cgg}
into our formalism by identifying $\psi=C$, and taking the
four-dimensional Einstein metric \beq g_{\mu\nu}(x)dx^\mu
dx^\nu=-e^{3(2B+nC)}dt^2+e^{2B+n C}d{\bf x}^2_{(3)}\,,
\eeq
which
describes a flat FLRW universe with scale factor \beq
a(t)=e^{B+nC/2}. \eeq Proper time (for a four-dimensional
observer) is measured by $\eta$, obtained from \beq d\eta=a(t)^3
dt\,, \eeq and the condition for acceleration is $d^2a/d\eta^2>0$.

The solutions are
\beq
B(t)=-\frac{1}{3}\log \left( b\sqrt{\frac{n-1}{6(n+2)}}\cosh
3(t-t_0)\right)\,,
\eeq
and
\beq
\psi(t)=C(t)=g(t)-\frac{3}{n-1}B(t)\,,
\eeq
where we have defined
\beqa
g(t) &=& \left\{ \begin{array}{ll}
 \frac1{n-1}\log\left(\frac{\beta}{\sinh[(n-1) \beta |t|
   ]} \right), \qquad & \sigma=-1, \\
  \beta t, & \sigma=0, \\
 \frac1{n-1}\log\left(\frac{\beta}{\cosh[(n-1) \beta t
   ]} \right), & \sigma=+1, \end{array} \right.
\eeqa
and
\beq
\beta=\frac{1}{n-1}\sqrt{\frac{3(n+2)}{n}}
\eeq
(for later use observe that $(n-1)\beta<3$ when $n>1$). We have
eliminated some of
the parameters in \cite{cgg} via coordinate redefinitions, and have also
chosen others so that we obtain the class of accelerating cosmologies.
The time shift $t_0$ does not make any qualitative difference when
$\sigma=0,-1$, but it is important for the spherical case $\sigma=+1$.
Note further that hyperbolic compactifications, $\sigma=-1$, present a
singularity at $t=0$. As in \cite{tw}, in this case we restrict
ourselves to $t<0$. Spacetime is future-complete since $t=0^-$
corresponds to $\eta\to\infty$. For $\sigma=0,+1$, $t$ runs from
$-\infty$ to $+\infty$. In all cases $t\to -\infty$ corresponds to
$\eta\to 0$, and this is interpreted as the big-bang singularity.

It is now a simple task to study the behavior of the volume scalar
$\psi$ in these solutions. For comparison with the extensive
literature on exponential potentials (see e.g. \cite{expcosmo} and
references therein), it is useful to write
\beq V(\psi)\sim
e^{-\lambda \hat\psi}, \label{lamb} \eeq where
$\hat\psi=[n(n+2)/2]^{1/2}\psi$ is the (dimensionless) canonically
normalized field. The nature of the attractors depends on whether
$\lambda^2>6$ (``steep potentials") or $\lambda^2<6$ (``flatter
potentials"). The flux term in (\ref{psipot}) is ``steep", $6 \leq
\lambda^2_{b} < 18$ and positive, whereas the curvature term is
``flatter", $2 < \lambda^2_{\sigma}\leq 4$, and can have either
sign.

At the big-bang, $t\to-\infty$, \beq \psi \to
\left(\frac{3}{n-1}-\beta\right)(-t)\to +\infty \,, \eeq for all
$b$ and $\sigma$, and \beq \frac{d\psi}{d\eta} \to
-\left(\frac{3}{n-1}-\beta\right) e^{\frac{3}{2}\left(n\beta
-\frac{n+2}{n-1}\right) (-t)} \to -\infty\,. \eeq Clearly, this is
a kinetic-dominated regime, with limiting $p=\rho$. This is an
unstable critical point for all types of exponential potentials,
so the subsequent evolution flows away from it.

\bigskip

For the evolution at later times we have to consider separately
each value of $\sigma$:

\bigskip

\noindent$\bullet$ For hyperbolic compactifications, $\sigma=-1$,
the asymptotic future is at $t\to 0^-$, and then \beq \psi\to
\frac{1}{n-1}\log \frac{1}{|t|} \to +\infty\,, \eeq so the field
rolls back down the hill, now with asymptotic velocity \beq
\frac{d\psi}{d\eta}\to |t|^{\frac{3n}{2(n-1)}}\frac{1}{n-1}\log
\frac{1}{|t|} \to 0^+\,. \eeq This corresponds to the stable
attractor of the potential $e^{-(n+2) \psi}$, where the potential
and kinetic energies are comparable. In the notation introduced
around Eq. (\ref{lamb}), such attractor exists only for flatter
potentials with $\lambda^2<6$. The effective equation of state in them
is given by $p=[(\lambda^2/3)-1]\rho$, where
$\lambda^2=\lambda^2_{\sigma}=2(n+2)/n$, and the scale factor
behaves as
\begin{equation}
a(\eta)\sim \eta^r \label{scf}\,,
\end{equation}
where $r=2/\lambda_\sigma^2=n/(n+2) < 1$. This is a slowly
decelerating, expanding universe. But, obviously, before reaching
this regime the field must have turned around, so that at some
finite value of $t<0$ the field halts, $\dot\psi=0$. This happens
when \beq \beta(n-1)\coth\beta(n-1)|t|=3 \tanh 3|t|\,. \eeq This
transcendental equation has a unique solution for $t<0$. At around
this instant we expect to find a phase of accelerated expansion.
This can be easily verified by plotting the evolution of $\psi$
together with the acceleration of the scale factor.

\bigskip

\noindent$\bullet$ For flat compactification, $\sigma=0$, we expect a
similar behavior of the accelerating phase, and this expectation is
indeed borne out by a closer examination. Asymptotic future is $t\to
+\infty$, and in this case $\psi \to (\beta +3/(n-1)) t\to\infty$. As
before, there is an instant of finite $t$ where $\dot\psi=0$, and around
it the expansion of the universe accelerates.

In contrast to the previous case, the asymptotic future evolution
is now kinetic-dominated. This is due to the fact that when
$\sigma=0$, the potential is controlled by the `flux term' in
(\ref{psipot}) $\sim e^{-3n\psi}$, which is steeper than the
internal volume factor $\sim e^{-(n+2)\psi}$, absent in this case.
For steep positive potentials ($\lambda^2>6$), the
kinetic-dominated expansion is precisely the expected late-time
attractor. So this is another example of how the main features of
the solutions can be anticipated easily by a direct examination of
the potential.

\bigskip

\noindent$\bullet$ Finally, for spherical compactifications,
$\sigma=+1$, the asymptotic future is different from the previous ones,
but in a way that could be anticipated from the attractor of the
negative exponential. For all values of the parameters, the above
solutions end up in a phase where $\psi\to\infty$ and the universe is
described by the collapsing attractor.

To see whether there is a phase of acceleration,
we need to take into account
the value of the parameter $t_0$. If we set it to zero, then the
solutions are time-symmetric around $t=0$. This is also the instant
where the field then turns around. At this moment,
\beq
\psi(t=0)=\frac{1}{n-1}\log \frac{b}{\sqrt{2n(n-1)}}\,.
\eeq
Precisely for this value of $\psi$ the potential vanishes, $V(\psi(t=0))=0$.
Hence there is never a
phase of positive potential energy domination, and we do not expect any
acceleration in this case. Examination of $d^2 a/d\eta^2$ shows this to
be the case. At this turning point the universe changes from
expansion to collapse.

However, if we allow for $t_0\neq 0$, the moment when expansion yields
to collapse is not an instant of time symmetry. Moreover, the field
overshoots and does reach positive values of $V$, so there is
acceleration. If $t_0>0$, then the acceleration occurs when the universe
is already collapsing, and acts for too short to be able to revert the
contraction. However, if $t_0<0$ the acceleration happens when the
universe is still expanding, so we obtain a brief phase of accelerated
expansion. These details are hard to discover from a simple examination
of the potential, but nevertheless they are at least easily understood
within this context.

During expansion, the motion of the field is damped by
cosmological friction. On the other hand, during a collapsing
phase, the motion of the field is anti-damped, and one might
wonder if this might not send the field up the flux-term potential
towards $\psi\to -\infty$, causing yet a different type of
behaviour without a turning point. It turns out, however, that
steep positive potentials with $\lambda^2> 6$ do not allow such
behaviour. Even with the help of anti-friction the field is unable
to climb indefinitely and has to turn around.

\bigskip

Our aim here has not been so much the detailed study of the
S-brane solutions of \cite{cgg}, but rather to stress that the
main features of the cosmologies that one can obtain is best
understood using quite conventional ideas of four-dimensional
cosmology.
In this context, it is worth recalling the possible uses of the
accelerating phases described above.

First of all, from solutions of the type given in \cite{tw}, it
appears quite difficult to achieve the sustained period of
acceleration needed in inflationary cosmology. An inflationary
attractor would require $r>1$ in Eq. (\ref{scf}), or
$\lambda^2<2$, too flat for the exponentials in (\ref{psipot}).

On the other hand, for applications to late time acceleration, where
$\psi$ might play the role of a quintessence field, the number of
e-foldings required is just of order one. However, the acceleration
found in the bounce period is not adequate for this purpose since we
know that our universe was not kinetic-dominated before entering
accelerated expansion. Finding an appropriate phase of acceleration with
the exponential terms in (\ref{psipot}) seems unlikely. The approach to
the attractor occurs during the radiation and matter dominated eras
\cite{ed}, and by the time the quintessence potential $V(\psi)$ becomes
cosmologically relevant the solution is well described by (\ref{scf}).
So again, acceleration would need $\lambda^2<2$. Furthermore, a
quintessential field must be very light, behaving as massless on scales
smaller than the Hubble radius. Although the radion $\psi$ is light at
large values, it cannot be used for quintessence, since it would mediate
long-range interactions of gravitational strength. These are subject to
strong observational constraints.

There are several obvious extensions. Compactification on an
Einstein manifold, with a reduction truncated to a single volume
scalar, leads to the same kind of four-dimensional Lagrangians we
have studied. One might also consider compactification on product
spaces. In these cases one might want to include a volume scalar
for each of the components. The resulting potential $V(\psi_i)$
would contain sums and products of exponentials, which may lead to
more useful cosmological behaviour. This is currently under
investigation.

\section*{Acknowledgements}
We are grateful to Paul Townsend for useful discussions. This work
is partially supported by grants UPV00172.310-14497, MCyT FPA2002-00748,
FPA2001-3598, DURSI 2001-SGR-00188 and
HPRN-CT-2000-00131.

\end{document}